# The Variable Multiple Bandpass Periodic Block Bootstrap for Time Series with Multiple Periodic Correlations

Edward Valachovic

evalachovic@albany.edu

Department of Epidemiology and Biostatistics, School of Public Health,

University at Albany, State University of New York, One University Place, Rensselaer,

NY 12144

**Abstract**

This work introduces a novel block bootstrap method for time series with multiple periodically correlated (MPC) components called the Variable Multiple Bandpass Periodic Block Bootstrap (VMBPBB). While past methodological advancements permitted bootstrapping time series to preserve certain correlations, and then periodically correlated (PC) structures, there does not appear to be adequate or efficient methods to bootstrap estimate the sampling distribution of estimators for MPC time series. Current methods that preserve the PC correlation structure resample the original time series, selecting block size to preserve one PC component frequency while simultaneously and unnecessarily resampling all frequencies. This destroys PC components at other frequencies. VMBPBB uses bandpass filters to separate each PC component, creating a set of PC component time series each composed principally of one component. VMBPBB then resamples each PC component time series, not the original MPC time series, with the respective block size preserving the correlation structure of each PC component. Finally, VMBPBB aggregates the PC component bootstraps to form a bootstrap of the MPC time series, successfully preserving all correlations. A simulation study across a wide range of different MPC component frequencies and signal-to-noise ratios is presented and reveals that VMBPBB almost universally outperforms existing methods that fail to bandpass filter the MPC time series.

# 1. Introduction

**1.1 Background**

There are many physical, environmental, biological, and human generated activities that exhibit periodically correlated (PC) patterns when measured and recorded across time, also known as time series data. These patterns may be related to natural activities such as the Earth's rotation exhibiting well known seasonal patterns exhibited in temperature, or human activities exhibiting weekly patterns in traffic congestion. A PC component in time series data is characterized by its particular correlation structure, where there is a strong correlation between observations separated by $p$ units of time. Time series definitions, notation, and examples can be found in Wei (1989). The period, $p$, or the corresponding reciprocal of the period, $1/p$, called the frequency, in addition to the periodic mean, or the mean value at each point of time throughout the period, are the characteristics that specify a given PC component. Additionally, many time series may be formed from multiple periodically correlated (MPC) components. Examples include those mentioned above, where daily temperature measured hourly may feature daily PC patterns driven by the Earth's rotation in addition to the annual, or seasonal, pattern resulting from the Earth's revolution about the sun. Similarly, hourly traffic may exhibit PC patterns in daily cycles caused by work arrival/departure times, weekly patterns resulting from the 7-day work week, and annual patterns due to holidays and seasonal activities. Furthermore, even a single PC component may not behave as a perfect sinusoidal wave, strictly at one principal period or frequency. If a PC component of a given period $p$, or frequency $1/p$ called the fundamental frequency, has a pattern other than a sinusoidal wave, it often has variations that change at integer multiples $j \in \mathbb{Z}$ of the fundamental frequency, $j \times (1/p)$, called the $j^{th}$ harmonic of the fundamental frequency. Therefore, a time series with only one PC component at a fundamental frequency may still be a MPC time series with the harmonics.

To better understand the characteristics of PC or MPC time series with a given period or periods, it is necessary to estimate their periodic means. Bootstrapping, or independent resampling from a given sample of data, can be a useful non-parametric approach to estimate the sampling distribution of statistical estimator such as the mean. Random sampling with replacement results in independent draws that form a resample of the same length as the original data, and replicates much of the sampling variability when used to calculate the distribution of possible values of a statistic and was first detailed by Efron (1979). In time series data, successive data points, or observations, are ordered. The ordering can be in space, space-time, or another combination of dimensions, generally described as spatio-temporal data, but without loss of generality the data can still be referred to as time series and the metric can be time. Clearly, independent random resamples of time series will not preserve the order of observation of a time series. This is a particular problem in the case where the ordered data points are correlated with prior observations in the time series. Consequently, independently resampling data points from the dataset to form a bootstrap also will not replicate correlations between given and prior data points.

A Block Bootstrap (BB) was introduced to preserve correlations in time series data. Block bootstrapping involves a general strategy of splitting the time series into blocks, and then independently resampling the blocks to form the bootstrap. An example of this is the Moving Block Bootstrap (MBB) first introduced by Kunsch (1989). This can help replicate the correlation structure between given data points and, at least up to a limited number, some prior observations. Correlations between data points that are separated by more than the length of the resampled blocks will likely not preserve correlation structures.

If a PC component has a given period, $p$, it exhibits a unique type of correlation structure with strong correlations between observations that are separated by $p$ time points. Some forms of resampling methods can adapt to maintain the order of time series, such as the permutation test from Pitman (1938). However, for the purposes of estimating the sampling distribution of the periodic mean, in general, many block bootstrapping methods have difficulty preserving the correlation structure of PC time series. First, arbitrarily bootstrapping with block lengths less than the given period, $p$, will fail to preserve these correlations. Additionally, bootstrapping with block lengths greater than the given period, $p$, will preserve the correlation within each block, however there is no reason that randomly resampled adjacent blocks forming the bootstrap will retain the periodic correlation structure between blocks. The block bootstrap design does not require that the first data point of a sampled block to be the next step in each cycle of period $p$ that follows the last data point of the prior sampled block. For this reason, many block bootstrapping strategies are not well suited for PC time series.

A block bootstrapping strategy that is better suited for PC time series with a PC component of period $p$, is one that partitions the sample into non-overlapping blocks of length equal to the period. The Seasonal Block Bootstrap (SBB) was proposed by Politis (2001). The SBB restricts the time series to be an integer multiple of the period, $p$. Regardless of the step in the cycle a block begins with, all other blocks in the resample will likewise start with a common step in the cycle of period $p$. This strategy will preserve correlation structures between data points that are $p$ points removed from each other in the bootstrap. Further developments advanced block bootstrapping to preserve a PC time series correlation structure including a procedure by Chan, Lahiri, and Meeker (2004) as well as the Generalized Seasonal Block Bootstrap (GSBB) by Dudek et al. (2014) and Generalized Seasonal Tapered Block Bootstrap (GSTBB) by Dudek, Paparoditis, and Politis (2016). While these methods bootstrap the original time series to preserve the PC component, they inadvertently bootstrap other interfering frequencies such as noise, which creates inefficiencies, loss of sensitivity, and increased bootstrap variability. Subsequently, to address that weakness common to those previous methods, collectively referred to here as periodic block bootstraps (PBB), the Variable Bandpass Periodic Block Bootstrap (VBPBB) was introduced by Valachovic (2024). Rather than bootstrapping the original time series, similar to PBB, this new method bandpass filters the original time series to reconstruct the PC component, and then bootstraps the PC component rather than the original time series, reducing interference from noise and other components. However, all of these strategies are similar in that given the period, $p$, of the PC time series component of interest, they preserve that one PC component in the bootstrap process by setting the block size equal to the period, $p$.

**1.2 Existing Methods and Challenge**

The PBB methods described above have some success preserving one PC correlation structure in time series by choosing block sizes equal to the period of the PC component. However, the reason the PBB approach is successful for one PC component, is precisely the reason it generally is not suitable for a MPC time series. The challenge presented by two or more MPC components contained within a time series is that choosing block size for one PC will almost certainly be unsuitable for another. Only in rare circumstances will the period of one PC component be an integer multiple of another. In this more trivial case, PBB methods can be performed for the longer period and still bootstrap the shorter without eliminating the correlation structure of either. Even so, this method will not properly resample the shorter period PC component since several cycles of the shorter PC component will be present within each block.

The more likely and more detrimental situation is when the MPC time series is composed of multiple PC components, each with their own correlation structure, where one periodicity will not be an integer multiple of another. As discussed above, PBB methods can preserve one PC component correlation structure but the choice of block length suitable for preserving that correlation structure will fail to preserve that of a different PC component. This is illustrated in Figure 1. Any choice of block length for resampling from correlated time series with two or more component periodicities will fail to preserve at least one inherent periodic component's correlation structure.

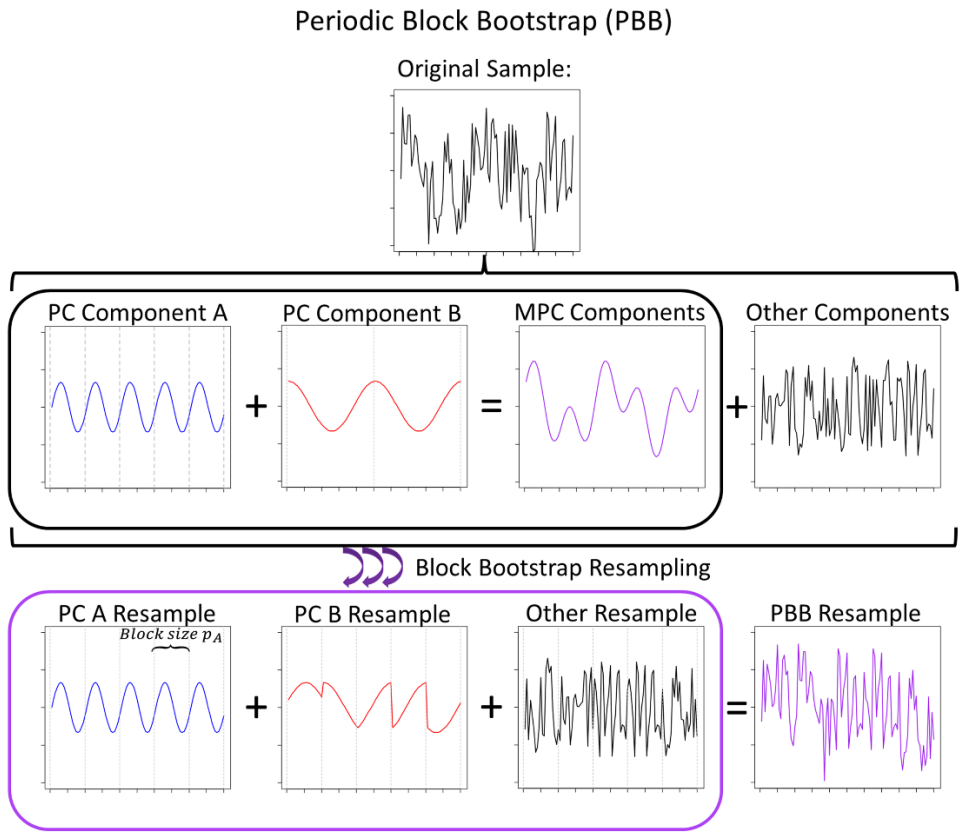

**Figure 1:** Illustration of PBB resampling a MPC time series composed of a PC component A of period $p_A$, a different PC component B of period $p_B$, and other components such as noise. The block bootstrapping strategy sets the block length to period $p_A$, preserving the

PC correlation structure of component A but fails to preserve the PC structure of component B, while resampling all components.

**1.3 The VMBPBB Solution for MPC Time Series**

The objective is to bootstrap MPC time series with two or more PC components while preserving the correlation structures of all periodic components present. One unwieldy solution for two components would be to choose a block length that is an integer multiple of the period of both components. In effect, the block length would be the least common multiple of the two periods. This however is problematic for several reasons. Unless the multiple is relatively small, and as a result the periods themselves are very small, the block length will be relatively large compared to the data set length. Also, this approach would fail to be a true resample of all of the PC components cycles. Furthermore, these problems are compounded if a MPC time series is composed of more than two PC components.

The proposed Variable Multiple Bandpass Periodic Block Bootstrap (VMBPBB) solution is inspired by the VBPBB from Valachovic (2024) and the periodogram of the time series, which is a spectral density or frequency domain representation of a time series and is described in Wei (1989). A mathematical fact is that two or more different PC components operating at different frequencies, may sum in the time domain but have zero correlation between them. Therefore, different frequencies operate independently. Separating and filtering a MPC time series according to its spectral density by the individual PC component frequencies, would create a set of component PC time series each with one PC correlation structure. Bandpass filters can accomplish this task. The set of component PC time series each with one PC structure could then individually be periodically block bootstrapped with appropriate block sizes to preserve the respective correlation structures. Aggregating the PC time series bootstraps will bootstrap the original MPC time series. This approach will preserve all PC component correlation structures within a MPC time series.

There are several distinct advantages in VMBPBB for MPC time series. The VMBPBB creates bootstrap resamples of each individual PC component as a step in the process to bootstrap the MPC time series. Therefore, VMBPBB is also useful if the primary purpose is to investigate the characteristics of each PC component individually, as any subset, or collectively as the full MPC time series. For example, preserving the correlation structure and bootstrapping just the annual PC component and all harmonics is possible with the

VMBPBB. VMBPBB can even be useful if the primary purpose is investigating non-PC components of a PC time series, where accurate estimation of the individual PC components or MPC components is necessary for removal as a nuisance component from a time series.

Another advantage is evident when comparisons are made to PBB. It might be reasoned that it would be possible to perform something like the VMBPBB using PBB but without bandpass filters, by block bootstrapping each PC component and then recombining to form bootstraps of subsets or the full MPC. However, such an approach with PBB suffers from a significant flaw. While the VMBPBB independently resamples the set of bandpass filtered PC component time series, each of which is limited to frequencies near the PC component principal frequency, the PBB resamples the entire original time series and the full spectrum of frequencies. By resampling all frequencies, including that from noise or other PC components known not to be correlated to the chosen PC component, the PBB will greatly increase the variability of the bootstraps. The PBB bootstrap resamples will not be representative of the sampling variability of just the intended PC component. Furthermore, repeating PBB for each PC component and then recombining to form bootstraps of a subset or the full MPC will only compound this design problem. Therefore, VMBPBB should outperform the PBB for MPC time series as well as for individual PC time series.

**1.4 The VMBPBB**

The VMBPBB approach relies upon filtering the original MPC time series to form the set of PC component time series and then block bootstrapping each of these with block sizes chosen to preserve their PC component correlation structure. There are a variety of filters that can be used in this process, and a variety of block bootstrapping methods such as those collectively referred to as PBB. Example filters include lowpass, highpass, and bandpass filters that alone or in combination can be designed so that the filter cut-offs, or the boundary between passed and attenuated bandwidth, sits between the PC component frequencies. This results in one PC component in each passed region of the spectra. Bandpass filters are the natural candidate when there are two or more PC components. As previously described, they can also be beneficial in the case of a single PC time series, as it will separate that PC component from other potential components, even those that are not PC such as noise. VMBPBB is designed with the Kolmogorov-Zurbenko (KZ) filter, and its extensions. These filters and extensions are described in Yang and Zurbenko (2010). KZ filters are a class of low-pass filters, but their extension, the Kolmogorov-Zurbenko Fourier Transform (KZFT) filter, is a bandpass filter, and in combination with difference filters, are a flexible way to produce low, high, and bandpass filters with fine control over filter cut-off frequencies. KZ filters are iterations of a simple moving average, providing for easy computation through algorithms. Also, the functional arguments of the filter provide a clear and direct explanation relating to the desired problem.

KZ filters and their extensions can separate portions of the frequency domain to exclude interfering frequencies. These filters have a history of use isolating frequencies in a variety of fields such as the environmental sciences, meteorology, and climatology. Examples include investigating ozone in Wise and Comrie (2005) and Gao et al. (2021), air quality in Kang et al. (2013) and Sezen et al. (2023), pollution in Zhang, Ma, and Kim (2018), and precipitation in De Jongh, Verhoest, and De Troch (2006). Many of these examples highlight the use of KZ filters to smooth data, reduce random variation, interpolate missing observations, and specifically separate and filter portions of the frequency domain prior to analysis. These features make KZ filters and their extensions ideal for use in VMBPBB.

After PC component separation, VMBPBB block bootstraps each individual PC component time series. As outlined earlier, using block lengths equal to the period of a particular PC component, VMBPBB will preserve each PC correlation structure. VMBPBB can potentially use a variety of different PBB methods when bootstrapping the individual PC component time series that results from the bandpass filters. Here, VMBPBB is designed with a periodic bootstrap approach similar to the GSBB of Dudek et al. (2014). For a given period, $p$, a time series of $n$ observations is block bootstrapped by creating $p$ exclusive and exhaustive subsets, each composed of one of each of the first $p$ observations and the integer multiples of that observation. Each resample is formed by independently randomly selecting from the $p$ subsets in sequence, repeated until $n$ observations are selected. VMBPBB then repeats this process for a large number, $B$, of resamples. Statistical estimators, such as the periodic mean are calculated for each resample, providing an estimate of the sampling distribution for that statistic of the PC component. Figure 2 illustrates the VMBPBB process of using multiple bandpass filters and independently resampling the resultant PC component time series to produce bootstraps of the PC components as well as aggregating results to form a bootstrap of the MPC components while preserving each PC component correlation structures.

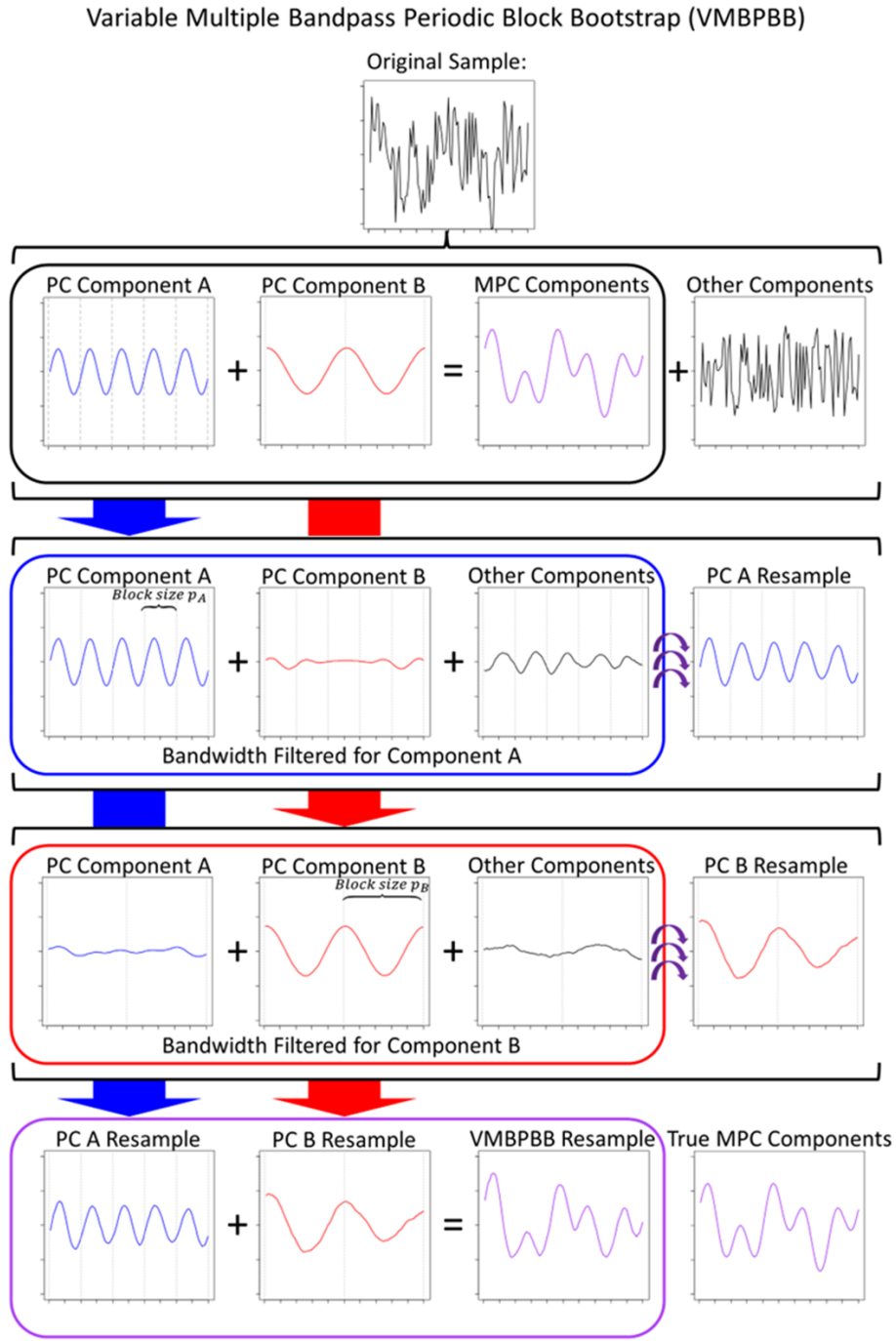

**Figure 2:** Illustration of VMBPBB resampling the MPC time series composed of a PC component A of period $p_A$, a different PC component B of period $p_B$, and other components

such as noise. The MPC time series is bandpass filtered prior to the block bootstrapping strategy that sets the block length to the period of the respective PC components, preserving the PC correlation structure of each component while limiting the resampling of other unintended components. Finally, bootstrapped PC components are recombined to form the MPC bootstrap.

## 2. Methods

### 2.1 The Bandpass Filter

The KZ filter is the iteration of a moving average of length *m*, a positive odd integer defined in Yang and Zurbenko (2010). It is a filter with two arguments, *m* is the filter window length and *k* the number of iterations. KZ filters are low pass filters that strongly attenuate signals of frequency 1/*m* and higher while passing lower frequencies. The equations below follow the notation found in Yang and Zurbenko (2010). Applied to a random time series process $\{X(t): t \in \mathbb{Z}\}$, a KZ filter of *m* time points and *k* iterations, where *t* is time, *u* are the steps defining the symmetric filter window, and $a_u^{m,k}$ are coefficient weights, is defined as follows:

$$KZ_{m,k}(X(t)) = \sum_{u=-\frac{k(m-1)}{2}}^{\frac{k(m-1)}{2}} \frac{a_u^{m,k}}{m^k} X(t+u) \qquad (2.1)$$

The coefficients $a_u^{m,k}$ can be conveniently found as the polynomial coefficients obtained from an expansion of the following polynomial (arbitrarily in *z*):

$$\sum_{r=0}^{k(m-1)} z^r a_{r-k(m-1)/2}^{m,k} = (1 + z + \cdots + z^{m-1})^k \qquad (2.2)$$

The KZ filter is more computational efficient when applied with statistical software in an iterated form using the KZA package in R software detailed in Close and Zurbenko (2013). The iterated form of the Kolmogorov-Zurbenko filter can be produced according to the following algorithm found in Yang and Zurbenko (2010) along with the following equations. In the iterated form of the $KZ_{m,k}$ filter, each step is an application of a $KZ_{m,1}$ filter to the prior result, equivalent to a moving average of size *m*, thus making all $a_u^{m,1}$ coefficients equal to one in each step:

$$KZ_{m,1}(X(t)) = \sum_{u=-\frac{m-1}{2}}^{\frac{m-1}{2}} \frac{a_u^{m,1}}{m^1} X(t+u) = \frac{1}{m} \sum_{u=-\frac{m-1}{2}}^{\frac{m-1}{2}} X(t+u) \qquad (2.3)$$

$$KZ_{m,2}(X(t)) = \frac{1}{m} \sum_{u=-\frac{(m-1)}{2}}^{\frac{(m-1)}{2}} KZ_{m,1}(X(t+u)) \qquad (2.4)$$

$$\vdots$$

$$KZ_{m,k}(X(t)) = \frac{1}{m} \sum_{u=-\frac{(m-1)}{2}}^{\frac{(m-1)}{2}} KZ_{m,k-1}(X(t+u)) \qquad (2.5)$$

Additionally, Yang and Zurbenko (2010) introduce the transfer function for the KZ filter, which is the linear mapping that describes how input frequencies are transferred to outputs. The energy transfer function is the square of the transfer function and as such is symmetric about zero. The energy transfer function of the KZ filter at frequency $\lambda$ is seen in the following equation. The energy transfer function is useful to show the effect of a KZ filter on a time series, and how with only a few iterations, this class of filters, strongly attenuates signals of frequency $1/m$ and higher while passing lower frequencies.

$$|B(\lambda)|^2 = \left(\frac{\sin(\pi m\lambda)}{m\sin(\pi\lambda)}\right)^{2k} \qquad (2.6)$$

The KZ energy transfer function is also useful to define the cut-off frequency which is a limit or boundary at which the energy transferred through a filter is suppressed or diminished rather than allowed to pass through. Control over both the energy transfer function and the cut-off frequency are useful in VMBPBB to ensure each filter applied to the MPC time series preserves only one PC component. A cut-off frequency where output power is half that of the input, called the half power point for the KZ filter transfer function is provided by Yang and Zurbenko (2010) below:

$$\lambda_0 \approx \frac{\sqrt{6}}{\pi} \sqrt{\frac{1-\left(\frac{1}{2}\right)^{\frac{1}{2k}}}{m^2-\left(\frac{1}{2}\right)^{\frac{1}{2k}}}} \qquad (2.7)$$

Where the KZ filter is symmetric around zero, the KZFT is a symmetric band pass filter equivalent to a KZ filter centered around frequency ν. Corresponding equations to those above for the KZFT filter and additional information follow and is found in in Yang and

Zurbenko (2010). KZFT is a filter applied to a random process $\{X(t): t \in T\}$ that has arguments $m$ time points, and $k$ iterations and center at a frequency $\nu$ and is defined:

$$KZFT_{m,k,\nu}(X(t)) = \sum_{u=-\frac{k(m-1)}{2}}^{\frac{k(m-1)}{2}} \frac{a_u^{m,k}}{m^k} e^{-i2m\nu u} X(t+u) \qquad (2.8)$$

The coefficients $a_u^{m,k}$ are the polynomial coefficients from equation (2.2).

Practical use of the KZFT filter is like the KZ filter since it can be produced in statistical software in Close and Zurbenko (2013). The energy transfer function of the KZFT filter at a frequency $\lambda$ with arguments $m$, $k$, and $\nu$ is given in the following equation:

$$|B(\lambda - \nu)|^2 = \left(\frac{\sin(\pi m(\lambda - \nu))}{m \sin(\pi(\lambda - \nu))}\right)^{2k} \qquad (2.9)$$

It follows that the cut-off frequency is:

$$|\lambda_0 - \nu| \approx \frac{\sqrt{6}}{\pi} \sqrt{\frac{1 - \left(\frac{1}{2}\right)^{\frac{1}{2k}}}{m^2 - \left(\frac{1}{2}\right)^{\frac{1}{2k}}}} \qquad (2.10)$$

For these filters, the cut-off frequency boundaries then become useful to determine the region of the spectra that is passed and that which is suppressed or filtered.

**2.2 KZFT Filter Selection**

To better understand the design of the KZFT filters that are used to separate MPC time series prior to block bootstrapping PC components, we consider the selection of each filter argument. Recall that the KZFT filter has three arguments, $m$, $k$, and $\nu$, which is the easiest to interpret since $\nu$ is the frequency around which the filter will be centered. The KZFT is a symmetric band pass filter. While the KZ filter is symmetric around the zero frequency, the KZFT is a KZ filter centered at argument $\nu$. Likewise, the KZFT filter reduces to a KZ filter when $\nu = 0$. For VMBPBB to bandpass filter each PC component separately from the MPC time series, VMBPBB can center a KZFT filter at the frequency corresponding to each PC component in the MPC time series. Therefore, for every PC component, one KZFT filter should have $\nu$ set to the PC component frequency.

The KZFT filter argument $m$ defines the width of the moving average filter window. It can be interpreted as defining the endpoints of the filter width, the frequency $1/m$ where the time series energy is completely attenuated. The effect of varying the *KZ* filter argument *m* is illustrated in Figure 3. Here we see that the argument *m* of the *KZ* filter shifts the frequencies at which the energy transfer is completely attenuated.

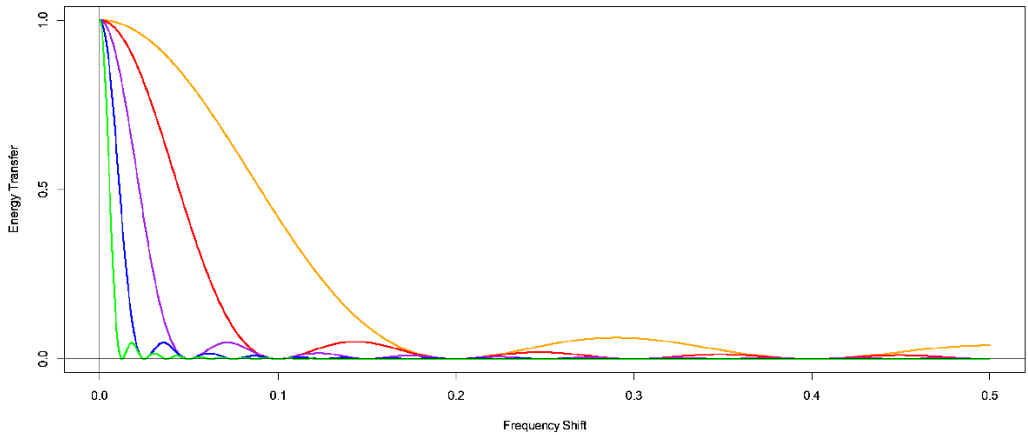

**Figure 3**: One side of the symmetric energy transfer function at different frequency shifts for a KZ filter centered at frequency zero, or equivalently a KZFT filter centered at frequency ν, with arguments $k = 1$ and $m = 5$ in orange, $m = 11$ in red, $m = 21$ in purple, $m = 41$ in blue, and $m = 81$ in green.

The KZFT filter argument *k* defines the number of iterations of the moving average filter that is performed. It can be interpreted as defining the sharpness of the edge of the filter, without altering the frequencies where the energy transfer is completely attenuated. The effect of varying the KZFT filter argument *k*, for a constant fixed *m*, is illustrated in Figure 4. Figure 4 shows that frequency that is completely attenuated does not change by changing only *k*, and the frequencies further away than $1/m$ are suppressed more strongly as *k* increases, unlike Figure 3.

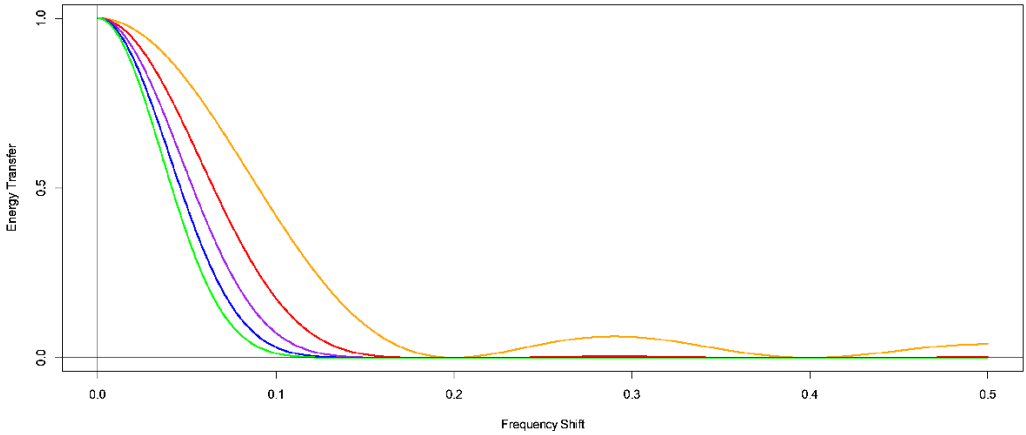

**Figure 4**: One side of the symmetric energy transfer function at different frequency shifts for a KZ filter centered at frequency zero, or equivalently a KZFT filter centered at frequency ν, with arguments $m = 5$ and $k = 1$ in orange, $k = 2$ in red, $k = 3$ in purple, $k = 4$ in blue, and $k = 5$ in green.

Figure 5 below illustrates a possible way two KZFT bandpass filters can separate one frequency band containing a specific frequency from another band containing the other frequency. For two PC components with periods $p_1$ and $p_2$ and corresponding frequencies $v_1 = 1/p_1$ and $v_2 = 1/p_2$, center the KZFT filters at $v_1$ and $v_2$. Set the arguments $k = 1$ and $m$ equal to the closest odd integer larger than $2/|v_1 - v_2| = 2(p_1 p_2)/|p_1 - p_2|$. This creates the bandpass KZFT filters in the figure. The frequencies where the energy transfer of each bandpass filter is completely attenuated are equal to the frequency midway between $v_1$ and $v_2$.

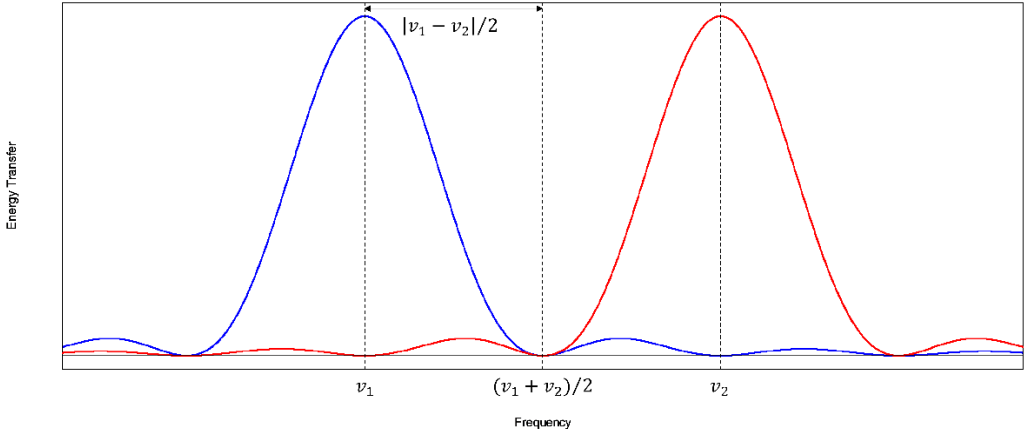

**Figure 5**: Illustration of two KZFT filters, one centered at frequency $v_1$ in blue and one centered at $v_2$ in red, with arguments $k = 1$ and $m$ equal to the closest odd integer larger than $2/|v_1 - v_2| = 2(p_1 p_2)/|p_1 - p_2|$. The blue bandpass contains frequency $v_1$ but excludes $v_2$, and the red bandpass contains frequency $v_2$ but excludes $v_1$.

The design described above provides one possible combination of arguments selected for the VMBPBB. For a MPC time series with $m$ PC components, where $i = 1, \ldots, m$, for each PC component with period $p_i$ and corresponding frequencies $v_i = 1/p_i$, the KZFT to bandpass that component has $v = v_i$, $k = 1$, and $m$ set to the largest value so that it excludes all other PC component frequencies as described above. The VMBPBB applies the first KZFT bandpass filter to the MPC time series, and supresses many other frequencies, particularly any other PC components, that lie outside of a narrow band around a first PC component frequency. This filter results in a time series that is composed principally of the first PC component only. This process is repeated for the second and subsequent PC components, filtering a narrow band around each frequency, excluding all others. This results in the set of PC component time series, each of which is principally composed of one and only one PC component correlation structure.

### 3. Simulations

The following simulation study tests the performance of VMBPBB under conditions and settings comparable to real world MPC time series data analysis. Furthermore, it compares VMBPBB to a similar approach, described previously, using PBB to bootstrap the individual PC components and collectively the entire MPC time series, but without bandpass filters under otherwise identical settings. These simulations support the VMBPBB theoretical advantages of bandpass filtering the original time series and block bootstrapping each resultant PC component time series. Without bandpass filtering the MPC time series, PBB bootstraps the original time series for each respective PC component and destroys some PC structures with each PBB application and unnecessarily increases variability of the bootstrap resamples.

**3.1 Simulation Methods**

Analysis is performed in R version 4.1.1 statistical software from the R Core Team (2013) using the KZFT function in the KZA package. More background on the KZA package is detailed in Close and Zurbenko (2013). All simulated time series are constructed with 1000-time units. In each simulation, two PC sine wave signals with different fixed periods, $p_1$ and $p_2$, are simulated with an amplitude of one unit producing the PC components. Next, random variation is introduced by generating equal length vectors of noise from a standard normal distribution with variance scaled to give the desired signal-to-noise ratio (SNR). These random variations are summed with the PC components to produce the MPC time series. An example of this simulation process is observed in Figure 6, where PC component 1 with period $p_1 = 50$ and PC component 2 with period $p_2 = 100$ and amplitudes of 1 unit, are summed with noise at a SNR of 1:10. This simulated MPC time series represents the data that ordinarily would be available at the time of analysis.

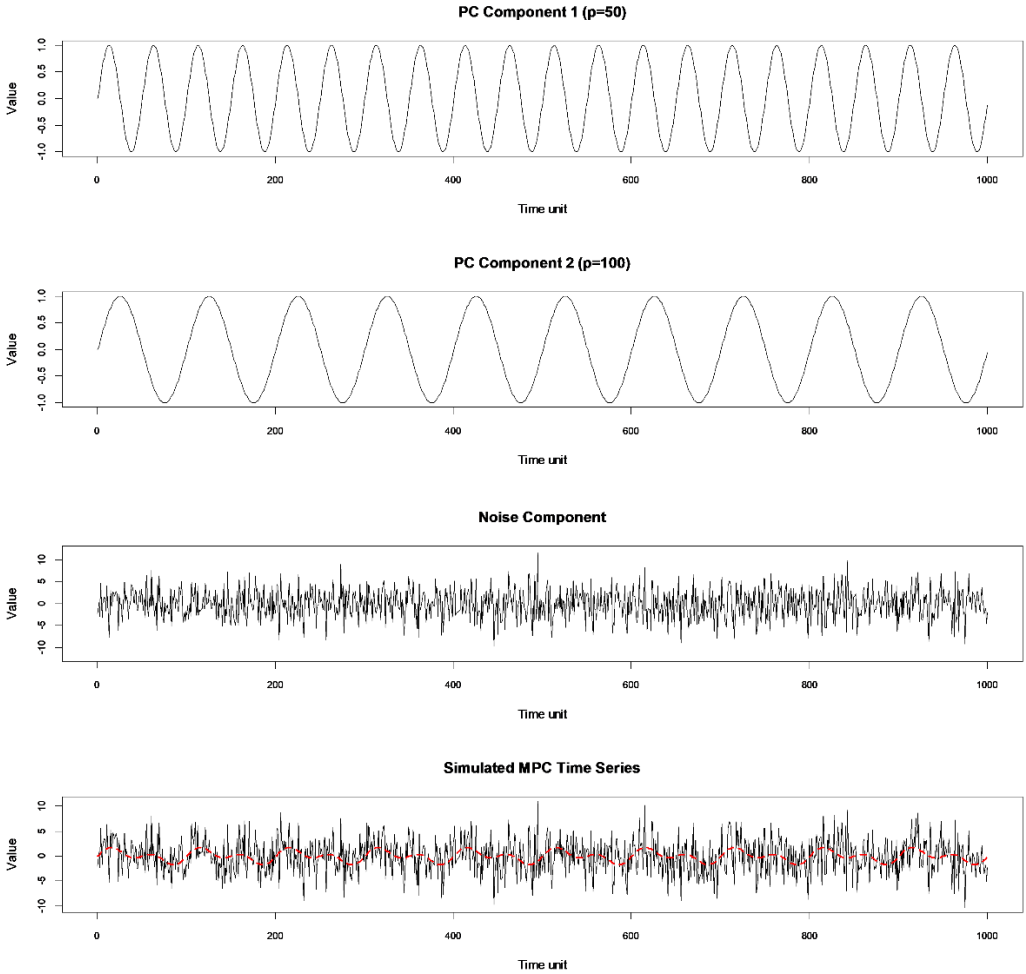

**Figure 6:** (From top) Simulated data of a PC component with $p_1 = 50$, PC component 2 with period $p_2 = 100$, noise component, and last the sum of all components in the simulated MPC time series in black with the sum of the MPC components in red.

That simulated MPC dataset is then used with the two approaches. The first approach applies the PBB without bandpass filtering the PC components. PBB block bootstraps the MPC time series to preserve each PC component correlation structure first by fixing block size for to $p_1$ and then $p_2$. The bootstraps for each PC component are then aggregated to create a bootstrap for the MPC, which preserves the correlation structure of both PC components. This will be referred to as the PBB approach. This is repeated for $B = 1000$ resamples. Next, on the identical dataset, VMBPBB separates the PC components using KZFT filters with arguments selected according to the procedure previously outlined. The R code used in this process can be found in Valachovic (2024). VMBPBB produces two

PC component time series, which are then block bootstrapped with their respective fixed block sizes $p_1$ and $p_2$. The bootstraps for each PC component are then aggregated to create a bootstrap for the MPC, which preserves the correlation structure of both PC components. This is also repeated for $B = 1000$ resamples. It is noteworthy that the PBB approach is identical to a special case of the VMPBB approach. Since the PBB approach resamples the original MPC time series, this is equivalent to a VMPBB approach where the bandpass filter trivially passes all frequencies, the same result when VMPBB uses KZFT filters with argument $m = 1$. Therefore, the comparison of the two approaches is directly interpretable as a test of the effect of bandpass filtering prior to bootstrapping a MPC time series.

To better evaluate performance, this simulation study examines a range of scenarios with different combinations of PC component periods and SNR. In each scenario, the $B = 1000$ bootstrap resamples are recorded and 95% bootstrapped confidence intervals (CI) are produced from the 0.975 and 0.025 quantiles for the mean at each time point. The mean at each time point may differ according to the accumulated effects of the MPC periodic components and is referred to as the MPC periodic mean. Therefore, a 95% CI band, which is the band across time formed from the individual 95% CI for the mean at each time point, is constructed for the MPC periodic mean value. To test consistency, in each scenario the production of the $B = 1000$ bootstrap resamples and accompanying CI bands are repeated 1000 times and results aggregated for the PBB and VMBPBB approaches. Finally, to test robustness, these simulations are performed within the scenarios of different PC periods including {10, 25, 50, 100, 250} to span a wide range of possible frequencies, differing order of PC component bandpass filtering although in theory this should have no effect, and at a SNR including {1:2, 1:5, 1:10} to represent a wide range of interference from other sources, such as noise. In total, across the MPC component scenarios including the three levels of SNR, five different periods, 1000 repetitions and $B = 1000$ bootstrap resamples in each repetition, this simulation study represents 60 thousand applications of bandpass filtering and 60 million simulated resamples through VMBPBB. Figure 7 provides a flowchart of the simulation study process and evaluation criteria.

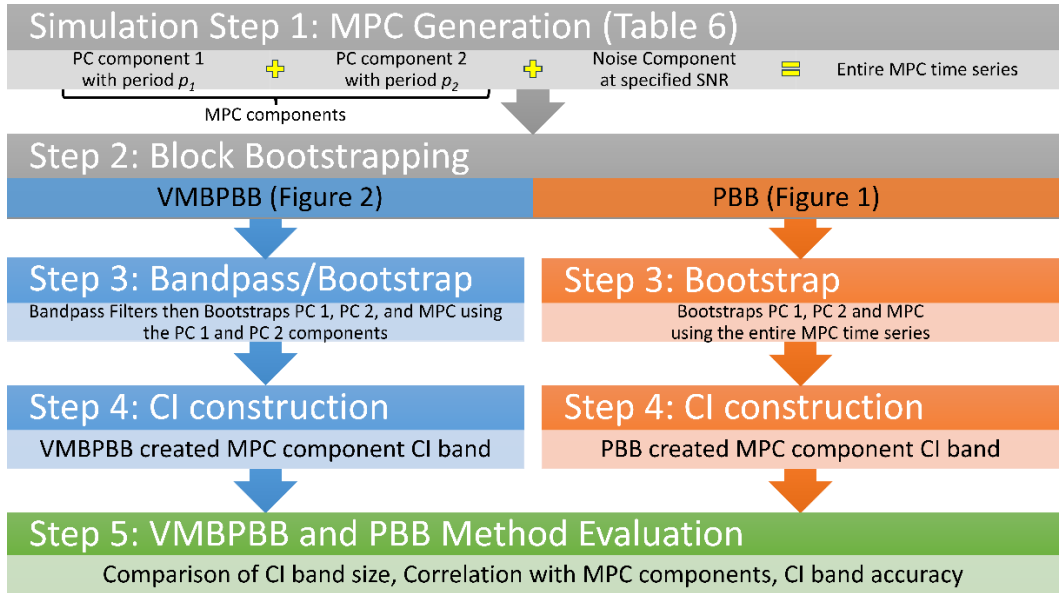

**Figure 7:** Simulation Flowchart.

**3.2 Assessment of Performance**

To assess the performance of the VMBPBB as compared to the PBB approach, we look at how well each method produces 95% CI Bands for the periodic mean for the MPC components, if those CI Bands are true to the 95% confidence level, and if the methods replicated the correlation structure of the MPC time series. First, at each combination of PC periods and SNR, and for each of the 1000 repetitions, the 95% CI band for the MPC periodic mean is created. The CI size difference along the CI bands is calculated comparing the PBB and VMBPBB approaches, and the median difference reported across the 1000 repetitions. In theory, we expect the CI band size to be smaller for VMBPBB. Additionally, in each scenario and for each repetition the proportion of the true MPC components lying within the 95% CI band is calculated, and then the median proportion is found among the 1000 repetitions. Finally, the median of the periodic means for the 1000 repetitions is calculated for the two approaches and the difference in the square of correlation with the original MPC components used in the simulation reported to see how well the correlation structures were preserved.

Figure 8 illustrates an example of one simulation scenario when the MPC time series is composed of a PC component with $p_1 = 50$, a PC component 2 with period $p_2 = 100$, and a SNR of 1:10. Performing PBB on the simulated MPC time series without bandpass filters, results in the 95% CI band for the periodic mean which spans the red region. While this region does contain the true periodic mean for the MPC components, represented in green, we notice this region representing 95% confidence is relatively wide and noisy with lower correlation with the MPC periodic mean of the true MPC components.

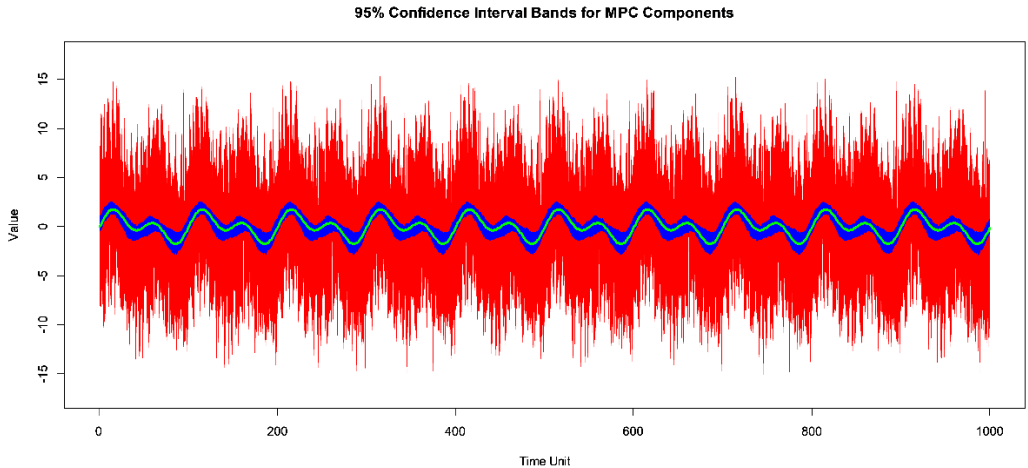

**Figure 8:** The MPC time series periodic mean, seen as the green line, the 95% CI band for the periodic mean of the MPC components using the PBB approach on the MPC time series without bandpass filtering in red and the 95% CI Band for the periodic mean of the MPC components by VMBPBB in blue.

If we contrast this with the blue region in Figure 8, representing the results of VMBPBB applied to the MPC time series by bandpass filtering and separately block bootstrapping the PC components, we observe relatively narrow and less noisy 95% CI bands for the MPC periodic mean. Here, the MPC periodic mean of the MPC components is contained within the blue 95% CI band but also shares higher correlation. VMBPBB has preserved both PC component correlation structures. While this is only the result from one illustrative example, to quantify the comparison we note that the PBB approach for the MPC time series in Figure 8 resulted in 95% CIs for the MPC periodic mean that at times ranged from approximately 5.61 times larger to 19.82 times larger than the CIs produced using VMBPBB. The PBB approach correlation squared with the true periodic mean of the MPC components was 38.7%, while VMBPBB was 95.4%, a difference of 56.7%. Finally, both PBB and VMBPBB CI bands contained the true MPC components at all simulated time points.

# 4. Results

For each combination of paired of MPC component periods and noise, Table 1 provides the median (across the 1000 scenario repetitions) ratio of 95% CI size comparing PBB to VMBPBB. The value represents the multiplicative factor, or how many times bigger, the 95% CI size is of the PBB approach as compared to the VMBPBB approach. The larger the CI, the worse the performance of the approach. A ratio of one indicates PBB and VMBPBB performed similarly, while the larger the ratio above one the better the performance of the VMBPBB approach. Across all simulated scenarios VMBPBB outperformed PBB. Across the scenarios of the simulation study design, ratios range 2.98 to 10.70 for a SNR of 1:2. Interpreting these values, this implies CIs produced by PBB are typically 2.98 to 10.7 times larger than those from VMBPBB. The ratios range from 2.86 to 12.30 for a SNR of 1:5, and 2.82 to 13.20 for a SNR of 1:10, demonstrating widespread superior performance by VMBPBB.

**Table 1**. Median ratio of 95% CI sizes for the MPC periodic means comparing PBB to VMBPBB.

| | 1:2 Signal-to-Noise Ratio (SNR) | | | | |
|---|---|---|---|---|---|
| | Period 2 (Frequency) | | | | |
| Period 1 (Frequency) | 10 (0.10) | 25 (0.04) | 50 (0.02) | 100 (0.01) | 250 (0.04) |
| 10 (0.10) | NA | 4.49 | 3.71 | 3.05 | 2.98 |
| 25 (0.04) | 4.45 | NA | 7.46 | 6.16 | 4.74 |
| 50 (0.02) | 3.69 | 7.49 | NA | 10.20 | 7.30 |
| 100 (0.01) | 3.04 | 6.18 | 10.10 | NA | 10.70 |
| 250 (0.04) | 2.99 | 4.76 | 7.41 | 10.60 | NA |
| | 1:5 Signal-to-Noise Ratio (SNR) | | | | |
| 10 (0.10) | NA | 4.22 | 3.61 | 2.99 | 2.86 |
| 25 (0.04) | 4.21 | NA | 7.47 | 6.20 | 4.72 |
| 50 (0.02) | 3.62 | 7.30 | NA | 10.50 | 7.78 |
| 100 (0.01) | 2.99 | 6.30 | 10.60 | NA | 12.30 |
| 250 (0.04) | 2.87 | 4.67 | 7.73 | 12.30 | NA |
| **1:10 Signal-to-Noise Ratio (SNR)** | | | | | |
| 10 (0.10) | NA | 4.09 | 3.57 | 2.98 | 2.82 |
| 25 (0.04) | 4.41 | NA | 7.12 | 6.19 | 4.75 |
| 50 (0.02) | 3.54 | 7.19 | NA | 10.10 | 7.64 |
| 100 (0.01) | 2.93 | 6.20 | 10.50 | NA | 13.20 |
| 250 (0.04) | 2.84 | 4.70 | 7.62 | 13.30 | NA |

For each combination of paired of MPC component periods and noise, Table 2 provides the median (across the 1000 scenario repetitions) difference in the square of correlation between the approaches' MPC periodic means and the true MPC component; for VMBPBB, less that from PBB. The value represents the additional percentage of the MPC component variance in the MPC periodic mean explained by the VMBPBB approach that is not explained by the PBB approach. The higher the square of correlation, or percent of variance explained, the better the approach performs in preserving the MPC correlation structure. A difference of zero indicates PBB and VMBPBB performed similarly, while the more positive the difference the better the performance of the VMBPBB approach. Excluding the scenarios when the paired periods are 10 and 25 for a SNR of 1:2 and 1:5, differences range from 12.88 to 47.70 for a SNR of 1:2, 20.93 to 68.81 for a SNR of 1:5, and 6.23 to 79.51 for a SNR of 1:10. This can be interpreted as VMBPBB explaining approximately 13 to 48% more of the MPC variability for a SNR of 1:2, 21 to 69% for a SNR of 1:5, and 6 to 80% for a SNR of 1:10 as compared to PBB again supporting widespread better performance by VMBPBB.

For the scenarios when the paired periods are 10 and 25 for a SNR of 1:2 and 1:5, seen as the entries with an asterisk on Table 2, with the argument selection design specified VMBPBB underperformed compared to PBB. This only occurred at this combination of periods and SNR of 1:2 and 1:5. However, recall that the KZFT filter argument $m$ is chosen so that the bandpass filter width is essentially half the distance between the PC components frequencies. In these scenarios, this may mean the bandpass filters are still too wide. In fact, with the small change of setting $m$ to approximately twice the original design, narrowing the bandpass regions, VMBPBB again outperformed PBB and has the differences provided in the table. This highlights of the need for flexibility of the arguments used for VMBPBB. This change came with the trade-off of CI size for the periodic mean, but with only a modest reduction in the advantage of VMBPBB. With the change, for scenarios where the periods are 10 and 25 and a SNR of 1:2, CI size for PBB was 3.44 times larger than from VMBPBB. For a SNR of 1:5, CI size for PBB was 3.29 times larger than from VMBPBB.

**Table 2**: Median difference of the square of correlation between the MPC periodic means and the true MPC component for VBPBB subtracting that from PBB.

| Period 1 (Frequency) | 1:2 Signal-to-Noise Ratio (SNR) | | | | |
|---|---|---|---|---|---|
| | Period 2 (Frequency) | | | | |
| | 10 (0.10) | 25 (0.04) | 50 (0.02) | 100 (0.01) | 250 (0.04) |
| 10 (0.10) | NA | 6.57* | 14.17 | 15.07 | 16.63 |
| 25 (0.04) | 8.68* | NA | 12.88 | 20.29 | 21.44 |
| 50 (0.02) | 14.96 | 13.64 | NA | 23.95 | 30.57 |
| 100 (0.01) | 14.76 | 20.96 | 24.23 | NA | 47.70 |
| 250 (0.04) | 16.82 | 21.54 | 32.22 | 48.13 | NA |
| **1:5 Signal-to-Noise Ratio (SNR)** | | | | | |
| 10 (0.10) | NA | 18.49* | 20.93 | 23.08 | 28.53 |
| 25 (0.04) | 14.57* | NA | 23.72 | 32.38 | 38.08 |
| 50 (0.02) | 21.08 | 23.31 | NA | 38.54 | 49.60 |
| 100 (0.01) | 23.00 | 31.21 | 37.96 | NA | 68.81 |
| 250 (0.04) | 28.14 | 36.84 | 48.55 | 69.42 | NA |
| **1:10 Signal-to-Noise Ratio (SNR)** | | | | | |
| 10 (0.10) | NA | 6.23 | 28.19 | 33.10 | 39.37 |
| 25 (0.04) | 6.46 | NA | 33.32 | 44.70 | 49.81 |
| 50 (0.02) | 27.56 | 34.46 | NA | 53.90 | 63.00 |
| 100 (0.01) | 33.05 | 44.15 | 51.28 | NA | 79.51 |
| 250 (0.04) | 37.62 | 50.52 | 63.49 | 79.36 | NA |

*Note*: * Asterisks indicate for these simulation scenarios VMBPBB used an argument $m$ approximately twice that of the standard design.

For each combination of paired of MPC component periods and noise, the median (across the 1000 scenario repetitions) difference in percentage of the periodic mean of the MPC components that falls outside a 95% CI band for the MPC periodic mean is calculated for VMBPBB less that from PBB. The lower the percentage of the PC component time series falling outside of bootstrapped 95% CIs for the periodic mean, the better. For a procedure producing 95% CIs, it would be anticipated, on average, approximately 5% of the periodic mean should fall outside of the interval. A difference of zero indicates PBB and VMBPBB performed similarly, while the larger the difference the better the performance of the PBB approach. Given the much larger CI sizes of PBB it is anticipated to perform better here, however, a VMBPBB performance similar to PBB would imply the much smaller CI size using VMBPBB and seen in Table 1 comes at little to no cost. All scenarios resulted in differences that were at or around approximately 0.05 or 5% or less, providing widespread evidence CI accuracy did not suffer under VMBPBB compared to PBB, and for this reason a table of differences is omitted.

## 5. Conclusion

The results of the simulations help compare the VMBPBB approach to that of PBB and provides evidence of the advantages of bandpass filtering a MPC time series as part of VMBPBB. Simulating across a wide range of scenarios of different constituent MPC component periodicities and with different SNR demonstrates the robustness of VMBPBB.

Furthermore, repeating each simulation 1000 times for every combination of PC periods and noise and then aggregating results demonstrates consistency of VMBPBB.

From the evidence provided by Table 1 within each scenario SNR, VMBPBB outperforms PBB. Closer inspection shows general trends that emerge in the performance of VBPBB. As the difference in paired PC component periods decreases, relative VMBPBB performance improves. It also improves dramatically for longer periods. Generally, as the SNR increases the relative performance improvement of VMBPBB verses PBB remains unchanged for shorter period PC components, but performance improves with increasing SNR for longer period PC components. These tendencies across the results may all be consequences of the argument selection design of the bandpass filters in this simulation study. Since the width of each band passed by the filters change, particularly with the choice of the $m$ argument as determine by the PC component periods, relatively more or less of other uncorrelated frequencies are retained affecting CI band size. It is worth noting that the argument selection design of the bandpass filters was created for simplicity across all the scenarios, and not with consideration in any sense to optimality. Still, even with this elementary argument choice, VMBPBB performance was near universally superior.

When examining the difference in the square of correlation between the approaches' MPC periodic means and the true MPC component, the evidence provided by Table 2 clearly demonstrates VMBPBB is performing better at preserving the MPC time series correlation structure across almost all scenarios. As identified in the results section, there were a few select scenarios where greater flexibility in the arguments used for VMPBPP was required to produce superior results to PBB. This highlights that in some cases a poor selection of arguments can degrade performance, and there is need for flexibility when selecting arguments for the bandpass filters.

Finally, the results demonstrated that for each combination of paired of MPC component periods and noise, the difference in percentage of the periodic mean of the MPC components that falls outside a 95% CI band for the MPC periodic mean between VMBPBB and that from PBB is minimal. This reinforces that VMBPBB is near universally able to produce often substantially smaller CI bands for MPC periodic means in MPC time series, with improved preservation of the true MPC component correlation structure, and with little or no impact on the accuracy of the constructed CI bands.

## 6. Discussion

Many natural and human-made processes exhibit PC and MPC correlation structures. Based on available information, there did not appear to be a suitable approach or method for bootstrapping a MPC time series that would preserve the correlation structures of all MPC components while estimating the sampling distribution of the MPC periodic mean. This work develops an approach and introduces the VMBPBB to accomplish this goal. Independently resampling the original MPC time series using available PBB methods to preserve one PC component is inefficient at best since other uncorrelated components at different frequencies are still resampled. When a time series has MPC components, performance further suffers, and problems are compounded. Any choice of block length for resampling directly from MPC time series with two or more PC component will likely destroy at least one PC component's correlation structure. Admittedly, PBB methods appear never to have been intended for MPC time series. However, for the purpose of comparison this work adapts the PBB into an approach to bootstrap MPC components collectively by aggregating bootstraps from each PC component.

This work then proposes the VMBPBB as a novel approach of bandpass filtering the MPC time series for each PC component, thereby creating a set of PC component time series that contain a narrow band of frequencies surrounding the fundamental frequency of the PC components. Each PC component time series contains one PC component from the original MPC time series. The PC component time series are then bootstrapped to preserve their respective correlation structures, and results are aggregated to form a bootstrap for the MPC time series, where all correlation structures remain. The VMBPBB method uses KZFT bandpass filters to create the PC component time series.

A simulation study across a wide range of PC component combinations, and with a variety of SNR demonstrates widespread superior performance by VMBPBB to PBB approaches for preserving correlation structures of MPC time series. VMBPBB bootstrapped CI band size for the MPC periodic mean were many times smaller than that of PBB, had higher, sometimes significantly higher, correlation with the true MPC components, and did not suffer significant loss in consistency with the designed CI size.

While the extensive simulation study within this work provides evidence of the success of the VMBPBB methodology, researchers may find it informative to see examples of practical applications to real-world problems, particularly in the use of the bandpass filtering and block bootstrapping programming code detailed in Valachovic (2024). For an example in the field of public health, significant MPC components were identified and investigated in US COVID-19 mortality in Valachovic and Shishova (2024). Yao and Valachovic (2024) use the new methodology to explore energy consumption with implications for conservation and economics. Finally, Sun and Valachovic (2024) use the VBPBB to investigate and identify multiple periodic correlations in particulate matter pollution. These recent examples demonstrate the practical use of the VMBPBB to real-world research as well as the wide applicability of this method.

There are limitations in this approach. This study illustrates the benefits of the VMBPBB approach, but it also shows importance of understanding the applications and limitations of KZFT filters. There are limitations to what frequencies are detectable, how closely two frequencies can be and still be separated by KZFT filters, and what data is required to investigate the separation of two or more specific frequencies. In addition, this research used VMBPBB with a rudimentary design for selecting KZFT filter arguments. While results showed this design in these scenarios was near universally successful, slight modification in the bandpass filter arguments was necessary to make it universally so. Additional work to determine optimal argument selection for VMBPBB is warranted. Also, the VMBPBB needs additional investigation to determine consistency under a variety of different factors including but not limited to relative PC component strength, additional frequency variation, additional PC components beyond two, non-PC components, and other levels of noise. Still, the results of this study seem conclusive, and the advantages of this new strategy are clear. The presence of multiple PC components prevents prior methods from accurately estimating the sampling distribution of MPC time series. The VMBPBB approach of separating PC components with bandpass filters prior to block bootstrapping the resulting individual PC component time series is necessary to preserve the correlation structures of multiple PC components in MPC time series.